\documentclass[12pt]{article}
\usepackage{amssymb,amsmath,epsfig}
\allowdisplaybreaks
\begin{document}

\title{\bf Stellar Filaments in Self-Interacting Brans-Dicke Gravity}

\author{M. Sharif $^1$ \thanks{msharif.math@pu.edu.pk} and Rubab
Manzoor $^{1,2}$
\thanks{rubab.manzoor@umt.edu.pk}\\
$^1$ Department of Mathematics, University of the Punjab,\\
Quaid-e-Azam Campus, Lahore-54590, Pakistan.
\\$^2$ Department of Mathematics,\\
University of Management and Technology,\\
Johar Town Campus, Lahore-54782, Pakistan.}
\date{}
\maketitle

\begin{abstract}
This paper is devoted to study cylindrically symmetric stellar
filaments in self-interacting Brans-Dicke gravity. For this purpose,
we construct polytropic filamentary models through generalized
Lane-Emden equation in Newtonian regime. The resulting models depend
upon the values of cosmological constant (due to scalar field) along
with polytropic index and represent a generalization of the
corresponding models in general relativity. We also investigate
fragmentation of filaments by exploring the radial oscillations
through stability analysis. This stability criteria depends only
upon the adiabatic index.
\end{abstract}

\section{Introduction}

Filamentary structures have important implications in structure
formation of the universe. These stellar configurations are
omnipresent in the universe at various scales. On cosmological
scales, cosmic filament is associated with the cosmic web \cite{2}
where it behaves like a bridge connecting various dense regions
(galaxies). On small scales, filaments are the features of
interstellar medium and instabilities within these filaments form a
dense medium which turns into a star \cite{3}. The stellar
filamentary structures appear in a variety of astronomical contexts,
their dynamical analysis can be considered as an effective tool to
understand the behavior of galactic as well as interstellar
structures in the universe.

It is very often that the geometry of the proposed models are taken
very simple or ideal partly because of reasons like mathematical
simplicity which may avoid complexities in the analysis, to obtain
some results of physical interest, to provide ideal model which can
act as a foundation and can be modified according to the required
astrophysical as well as cosmological applications. Cylindrically
symmetric configuration is one of the ideal case which has widely
been used to represent filamentary structure in the universe. In
1953, Chandrasekhar and Fermi \cite{4} described the dynamics of
cylinder filled with homogenous and incompressible fluid.
Stodolkiewicz \cite{5} developed the magneto-hydrodynamics
equilibrium of isothermal cylindrical filaments. Ostriker \cite{6}
studied cylindrically symmetric filamentary structure with a
polytropic equation of state and generalized the dynamical analysis
of homogenous compressible filaments. Afterwards, many researchers
explored cylindrical filamentary structures both analytically as
well as numerically \cite{8}. Recently, Breysse et al. \cite{9}
described polytropic filamentary structures through stability
analysis of cylindrically symmetric self-gravitating fluid.

Modified theories of gravity constructed by modifying the
Einstein-Hilbert action are considered to be candidates for
accelerating agent (dark energy) of the expanding universe
\cite{10}. There is a large body of literature \cite{11} which has
explored dynamics of stellar structures in modified gravity to
reveal the modification hidden in the structure formation of the
universe. Brans-Dicke (BD) gravity being a natural generalization of
general relativity (GR) \cite{13} is one of the fascinated example
of modified gravity which is considered as a solution of various
cosmic problems. One of the main features of this theory is that the
gravitational force is described by a massless scalar field $\phi$
along with curvature part (Ricci scalar). It also contains a
coupling constant $\omega_{BD}$ which serves as a tuneable parameter
to adjust the required results. This theory provides suitable
solutions of many cosmic issues but remains unable to describe
``graceful exist'' problem of old inflationary cosmology. The
inflationary model of this theory is valid for a specific value of
the coupling parameter $\omega_{BD}\leq25$ \cite{14} which
contradicts observational data \cite{15}. Moreover, the
observational limit of $\omega_{BD}$ at small scale (weak field)
\cite{16} is inconsistent with those calculated at large scale
\cite{17}.

These issues was resolved in self-interacting BD (SBD) gravity which
is developed by introducing a massive potential function $V(\phi)$
in the Jordan framework of BD gravity \cite{18}. The potential
function represents a massive scalar field term which describes the
potential of the scalar field $(\phi)$ or potential of energy
density related to scalar field in the system. This theory provides
a consistency between weak field as well as strong field regime
\cite{19}.  Recent literature indicates dynamics of SBD gravity in
many cosmic problems \cite{20}. In recent papers \cite{21}, we have
explored spherically as well as cylindrically symmetric
self-gravitating fluids in SBD gravity and found some interesting
results. We have also investigated hydrodynamics and oscillations of
spherically symmetric gaseous distributions in post-Newtonian
approximations of this theory \cite{22}.

In this paper, we study filamentary structures of stellar systems in
SBD theory. We construct cylindrical polytropic models of filaments
in the Newtonian (N) approximation and explore fragmentation of
filamentary structures through radial oscillations of
self-gravitating fluids. The paper is organized as follows. Section
\textbf{2} formulates SBD as well as dynamical equations in N limit.
In section \textbf{3}, we construct cylindrical polytropic
filamentary models in N approximation. Section \textbf{4} is devoted
to discuss radial oscillations of filaments through stability
analysis. Finally, section \textbf{5} summarizes the results.

\section{Self-Interacting Brans-Dicke Gravity and Dynamical Equations}

The action of SBD gravity \cite{18} is given by
\begin{equation}\label{1}
S=\frac{1}{2\kappa^{2}}\int d^{4}x\sqrt{-g} [\phi
R-\frac{\omega_{BD}}{\phi}\nabla^{\alpha}{\phi}\nabla_{\alpha}{\phi}-V(\phi)]
+L_{m},
\end{equation}
where $\kappa^{2}=\frac{8\pi G}{c^{2}}$ and $L_{m}$ shows matter
contribution. The variation of this action with respect to scalar
field ($\phi$) and metric tensor ($g_{\mu\nu}$) provides SBD
equations as follows
\begin{eqnarray}\nonumber
G_{\mu\nu}&=&\frac{\kappa^{2}}{\phi}T_{\mu\nu}+\frac{1}{\phi}[\phi_{,\mu;\nu}
-g_{\mu\nu}\Box\phi]+\frac{\omega_{BD}}{\phi^2}[\phi_{,\mu}\phi_{,\nu}
-\frac{1}{2}g_{\mu\nu}\phi_{,\alpha}\phi^{,\alpha}]
-\frac{V(\phi)}{2\phi}g_{\mu\nu},\\\label{2}
\\\label{3}
\Box\phi&=&\frac{\kappa^{2}T}{3+2\omega_{BD}}
+\frac{1}{3+2\omega_{BD}}[\phi\frac{dV(\phi)}{d\phi}-2V(\phi)],
\end{eqnarray}
where $T_{\mu\nu}$ represents the energy-momentum tensor of matter,
$T=g^{\mu\nu}T_{\mu\nu}$ and $\Box$ shows the d'Alembertian
operator. Equations (\ref{2}) and (\ref{3}) provide the field
equations and evolution of the scalar field, respectively. We assume
matter contribution in the form of a perfect fluid which can be
compatible with N regime
\begin{equation}\label{4}
T_{\mu\nu}=(\rho c^2 +p)u_{\mu}u_{\nu}-pg_{\mu\nu},
\end{equation}
where $\rho,~p,~u_{\mu}$ stand for density, pressure and four
velocity, respectively.

\subsection{Newtonian Approximation}

The weak field approximated solutions of any relativistic theory
describe the order of deviations or perturbation of the local system
from its homogenous and isotropic vacuum background. The N and
parameterized post-Newtonian limits are widely used weak field
approximated solutions that are derived by Taylor expansion of
metric functions as follows
\begin{eqnarray}\nonumber
g_{\mu\nu}&\approx&\eta_{\mu\nu}+h_{\mu\nu},\\\nonumber
\end{eqnarray}
with
\begin{eqnarray}\nonumber
h_{00}&\approx& h^{(2)}_{00}+h^{(4)}_{00}+\ldots,\\\nonumber
h_{0i}&\approx& h^{(3)}_{0i}+h^{(5)}_{0i}+\ldots,\\\nonumber
h_{ij}&\approx& h^{(2)}_{ij}+h^{(4)}_{ij}\ldots,
\end{eqnarray}
Here $\eta_{\mu\nu}$ indicates the Minkowski metric (describing
homogenous and isotropic vacuum background of $g_{\mu\nu}$),
$h_{\mu\nu}$ shows deviation of $g_{\mu\nu}$ from its background
values $(\eta_{\mu\nu})$, $i,j=1,2,3$ and the superscripts
$(2),~(3)$ and $(4)$ represent up to order of approximation
$(c^{-2}),~(c^{-3})$ as well as $(c^{-4})$. The N limits require the
information of $g_{00}\sim
\eta_{00}+h^{(2)}_{00},~g_{ij}\sim\eta_{ij}$ whereas parameterize
post-Newtonian corrections uses approximations $g_{00}\sim\eta_{00}+
h^{(2)}_{00}+h^{(4)}_{00},~g_{0i}\sim h^{(3)}_{oi}$ and $g_{ij}\sim
\eta_{ij}+h^{(2)}_{ij}$. Thus, the N limits of any system can be
directly obtained from its known parameterized post-Newtonian
approximations.

In order to discuss polytropic geometry in SBD gravity and check
compatibility of our results with the analysis of GR \cite{9}, we
approximate the system in N limits. For this purpose, we evaluate N
limits of SBD gravity from its known post-Newtonian approximated
solutions. The parameterized post-Newtonian approximation of SBD
(massive BD gravity) solutions has been evaluated by using the
following Taylor expansion of metric and dynamical scalar field
\cite{23}
\begin{eqnarray}\nonumber
g_{\mu\nu}&\approx&\eta_{\mu\nu}+h_{\mu\nu},\\\nonumber
\phi&\approx&\phi_{0}(t_{0})+\varphi^{(2)}(t,x)+\varphi^{(4)}(t,x),\\\nonumber
V(\phi)&\approx& V_{0}+\varphi V'_{0}+\varphi^{2}V''_{0}/2+....
\end{eqnarray}
Here $\phi_{0}$ describes dynamical scalar field as a function of
background cosmic time $t_{0}$ (that varies slowly with respect to
cosmic time $t_{0}$), $V_{0}=V(\phi_{0})$ is the potential function
of scalar field at $t_{0}$, and $\varphi(t,x)$ represents local
deviation of scalar field from $\phi_{0}$. The lowest-order
parameterized post-Newtonian corrections $(O(c^{-2}))$ of SBD
solutions are given by
\begin{eqnarray}\label{a}
g_{00}&\approx& 1-h^{(2)}_{00}=1-\frac{2U}{c^{2}}
+\frac{V_{0}r^{2}}{6\phi_{0}c^{2}},\\\label{b} g_{ij}&\approx&
-[1+h^{(2)}_{ij}]\delta_{ij}=[-1-\frac{2\gamma_{BD}
U}{c^{2}}-\frac{V_{0}r^{2}}{6\phi_{0}c^{2}}]\delta_{ij},\\\label{c}
\frac{\varphi^{(2)}}{\phi_{0}}&\approx&\frac{-2U}{c^{2}}
\left[\frac{e^{-m_{0}r}}{3+2\omega_{BD}+e^{-m_{0}r}}\right].
\end{eqnarray}
Here $U=G_{eff}\frac{M_{\odot}}{r}$ ($M_{\odot}$ is the Newtonian
mass of the sun) shows the effective gravitational potential
determined by the Poisson's equation
\begin{equation}\label{d}
\nabla^{2}U=-4\Pi\rho G_{eff},
\end{equation}
where
\begin{equation}\label{d'}
G_{eff}=\frac{\kappa^{2}}{8\pi\phi_{0}}\left(1
+\frac{e^{-m_{0}r}}{3+2\omega_{BD}}\right),\quad
m_{0}=\left(\frac{\phi_{0}V''_{0}-V'_{0}}{3+2\omega_{BD}}\right)^{1/2}.
\end{equation}

The term $\gamma_{BD}$ represents the parameterized post-Newtonian
parameter given by
\begin{equation}\nonumber
\gamma_{BD}=\frac{3+2\omega_{BD}-e^{-m_{0}r}}{3+2\omega_{BD}+e^{-m_{0}r}},
\end{equation}
where $m_{0}$ represents mass of the massive scalar field with
constraint $(m_{0}>>\frac{1}{\tilde{r}})$ (where $\tilde{r}$
represents scale of the experiment or observation testing the
field). When the background value of this mass is very small
$(m_{0}<<\frac{1}{\tilde{r}})$ the SBD system reduces to simple BD
gravity (massive scalar field reduces to a massless scalar field)
having post-Newtonian parameter
\begin{equation}\nonumber
\gamma_{BD}=\frac{1+\omega_{BD}}{2+\omega_{BD}}.
\end{equation}
That is why the BD theory (massless scalar field) is consistent with
solar system constraints of the Cassini mission for
$\omega_{BD}>40000$. However, for massive BD gravity (SBD gravity),
dynamics of the spatial part of $\phi$ is frozen on the solar system
scales through potential function and all values of $\omega_{BD}$
are observationally acceptable \cite{a}.

The term
$\frac{V_{0}r^{2}}{6\phi_{0}c^{2}}=\frac{\Lambda_{BD}r^{2}}{3c^{2}}$
describes the cosmological constant term which depends upon the
potential of the scalar field. In order to be consistent with
observational data ranging from the solar system to cluster of
galaxies the contribution due to scalar density should be very small
and the following constraint must be satisfy
\begin{equation}\nonumber
\frac{V_{0}L^{2}}{\phi_{0}}<<1.
\end{equation}
Here $L$ indicates length scale equal to or greater than solar
system.

The N limits of any scalar-tensor theory required information of
$g_{00}\sim\eta_{00}+h^{(2)}_{00},~g_{ij}\sim \delta_{ij}$ and
$\phi\sim\phi_{0}$. Thus, from the parameterized post-Newtonian
analysis of SBD theory the obtained N approximated SBD solutions are
evaluated as
\begin{eqnarray}\nonumber
g_{00}\approx 1-h^{(2)}_{00}=1-\frac{2U}{c^{2}}
+\frac{\Lambda_{BD}r^{2}}{3c^{2}},\quad g_{ij}\approx
-\delta_{ij},~\phi\approx\phi_{0}.
\end{eqnarray}
From now on, we use $h^{(2)}_{00}=h_{00}$ for the sake of
convenance. It can be noticed that by applying the limits
$(m_{0}<<\frac{1}{\tilde{r}})$ and
$\frac{V_{0}}{\phi_{0}}\rightarrow0$, the above defined N
approximations of SBD gravity can be shifted to N limits of BD
solutions. Similarly, in the limits
$(m_{0}<<\frac{1}{\tilde{r}})~,\frac{V_{0}}{\phi_{0}}\rightarrow0$
and $\omega_{BD}\rightarrow\infty$, the obtained approximated system
can be converted into GR case.

\subsection{Dynamical Equations}

The polytropic geometry of any configuration is base upon two
dynamical equations, the Poisson equation and the equation of motion
of the respective system \cite{9}. Here, we calculate the
generalized form of both dynamical equations for SBD gravity.
Equation (\ref{2}) can be rewritten as
\begin{eqnarray}\nonumber
R_{\mu\nu}=\frac{\kappa^{2}}{\phi}(T_{\mu\nu}-\frac{1}{2}g_{\mu\nu}T)
+\frac{\omega_{BD}}{\phi^{2}}[\phi_{,\mu}\phi_{,\nu}]+\frac{1}{\phi}[\phi_{,\mu;\nu}]
+\frac{g_{\mu\nu}}{2\phi}[\Box\phi+V(\phi)].
\end{eqnarray}
The temporal component of this equation in N approximation is given
by
\begin{equation}\label{8'}
R_{00}=\frac{1}{2}(\frac{k^{2}\rho}{\phi_{0}}-\frac{V_{0}}{\phi_{0}}),
\end{equation}
where the contribution due to $\dot{\phi_{0}}$ and $\ddot{\phi_{0}}$
are neglected because the term $\phi_{0}$ behaves almost constant.
The expansion of Ricci tensor in N regime is approximated as
\begin{equation}\label{8''}\
R_{00}=-\frac{1}{2}\nabla^{2}h_{00}.
\end{equation}
Comparing these two values, we obtain the generalized Poisson
equation as follows
\begin{equation}\label{7}
\nabla^{2}h_{00}=-\frac{k^{2}\rho}{\phi_{0}}+\frac{V_{0}}{\phi_{0}},
\end{equation}
where $h_{00}$ represents gravitational potential due to matter as
well as massive scalar field. The generalized Euler equations can be
obtained by using  $T^{\mu\nu}_{~~;\nu}=0$, whose time component
yields continuity equation
\begin{eqnarray}\label{8} \frac{\partial \rho}{\partial
t}+\frac{\partial}{\partial x_{i}}\left(\rho v_{i}\right)=0,
\end{eqnarray}
where $v_{i}~(i=1,2,3)$ indicate components of velocity. The spatial
components give the generalized form of equation of motion as
follows
\begin{eqnarray}\label{9}
\rho\frac{\partial v_{i}}{\partial t}=-\frac{\partial p}{\partial
x_{i}} -(\rho c^{2}+p)\frac{\partial(\ln g_{00})}{\partial x_{i}}.
\end{eqnarray}

\section{Cylindrical Polytropes}

In order to discuss cylindrical polytropic filaments in SBD gravity,
we consider equilibrium configuration and standard polytropic
equation of state as
\begin{equation}\label{10}
p=K\rho^{\gamma},
\end{equation}
where $K$ is a constant and $\gamma=\frac{n+1}{n}$ represents
polytropic exponent with $n$ as a polytropic index \cite{9, b}. In
hydrostatic equilibrium, the cylindrical configurations of
Eqs.(\ref{7}) and (\ref{9}) turn out to be
\begin{eqnarray}\label{11}
\frac{1}{r}\frac{d}{dr}\left(r\frac{dh_{00}}{dr}\right)
=-\frac{k^{2}\rho}{\phi_{0}}+\frac{V_{0}}{\phi_{0}},\\\label{12}
\frac{d p}{dr}=-(\rho c^{2}+p)\frac{d (\ln g_{00})}{dr}.
\end{eqnarray}
Integration of Eq.(\ref{12}) by using (\ref{10}) provides
\begin{equation}\label{11a}
\rho=\left[\frac{1}{K}\left(g_{00}^{-\frac{1}{2(n+1)}}
-1\right)\right]^{n}.
\end{equation}
Using binomial expansion on $g_{00}^{-\frac{1}{2(n+1)}}$ and
approximating upto $O(c^{-2})$, we have
\begin{equation}\label{11b}
\left(g_{00}^{-\frac{1}{2(n+1)}}
-1\right)\approx\frac{-1}{2(n+1)}h_{00}.
\end{equation}

Using Eq.(\ref{11b}) into Eq.(\ref{11a}), we obtain the relation
between density and potential configurations as follows
\begin{equation}\label{13}
\rho=C_{n}(-h_{00})^{n},\quad C_{n}=\left[2K(n+1)\right]^{-n}.
\end{equation}
Equations (\ref{11}), (\ref{13}) and cosmological constant term
provide a differential equation for a gravitational potential
$h_{00}$ as follows
\begin{equation}\label{14}
\frac{1}{r}\frac{d}{dr}\left(r\frac{dh_{00}}{dr}\right)
=-\frac{k^{2}C_{n}(-h_{00})^{n}}{\phi_{0}}+2\Lambda_{BD}.
\end{equation}
The above equation can be converted to polytropic equation
(Lane-Emden equation) by considering the following dimensionless
variables
\begin{equation}\label{15}
\beta(s)=\frac{h_{00}}{h_{00(c)}}=\left(\frac{\rho}{\rho_{c}}\right)^{\frac{1}{n}},\quad
s=\frac{r}{b},\quad
b=\left[C_{n}h_{00(c)}^{n-1}\right]^{-\frac{1}{2}},
\end{equation}
where $h_{00(c)}$ and $\rho_{c}$ represent potential as well as
density at the center of cylinder ($r=0$) \cite{9,b}. With these
assumptions, Eq.(\ref{14}) turns out to be
\begin{equation}\label{16}
\frac{1}{s}\frac{d}{ds}\left[s\beta\right]=-\beta^{n}+A,\quad
A=\frac{2\Lambda_{BD}}{h_{00(c)}}.
\end{equation}

This is a modified form of the original Lane-Emden equation
describing spherical polytropic model \cite{b} and generalized form
of modified Lane-Emden equation that describes cylindrically
symmetric polytropic filament in GR \cite{9}. Here $\beta(s)$ is the
Lane-Emden function describing equilibrium potential and $A$ is
considered as cosmological constant term due to the presence of
scalar field. This is a second order differential equation
satisfying the following boundary conditions: at the center ($r=0$)
we have $s=0,~h_{00}\approx
h_{00(c)},~\rho\approx\rho_{c},~\beta\approx1$ and
$\left(\frac{d\beta}{ds}\right)_{s=0}=0$. The surface of the
polytopic filament is represented by a specific value $s=S$ for
which density $\rho$ as well as the Lane-Emden function $(\beta)$
becomes zero. Equation (\ref{16}) can be solved analytically for
different values of $n$. Some possible analytical solutions are
given as follows
\begin{eqnarray}\label{17}
\beta(s)&=&1+\frac{s^{4}}{4}+\frac{As^{2}}{4},\quad n=0,\\\label{18}
\beta(s)&=&-A+J_{0}(s)+AJ_{0}(s),\quad n=1,
\end{eqnarray}
where $J_{0}(s)$ is a Bessel function of zeroth order. It is noted
that the analytical results depend upon the values of $s$ as well as
$A$. We can also solve Eq.(\ref{16}) numerically for different
values of $n$ and $A$. Some numerical solutions of the modified
Lane-Emden equation for $n=0,1,3,5$ with $A=0,\pm1/2,\pm1,\pm3$ are
given in Figures \textbf{1a}, \textbf{1b}, \textbf{2a}, \textbf{2b},
\textbf{3a}, \textbf{3b}, \textbf{4a} and \textbf{4b}, respectively.
\begin{figure}\centering \epsfig{file=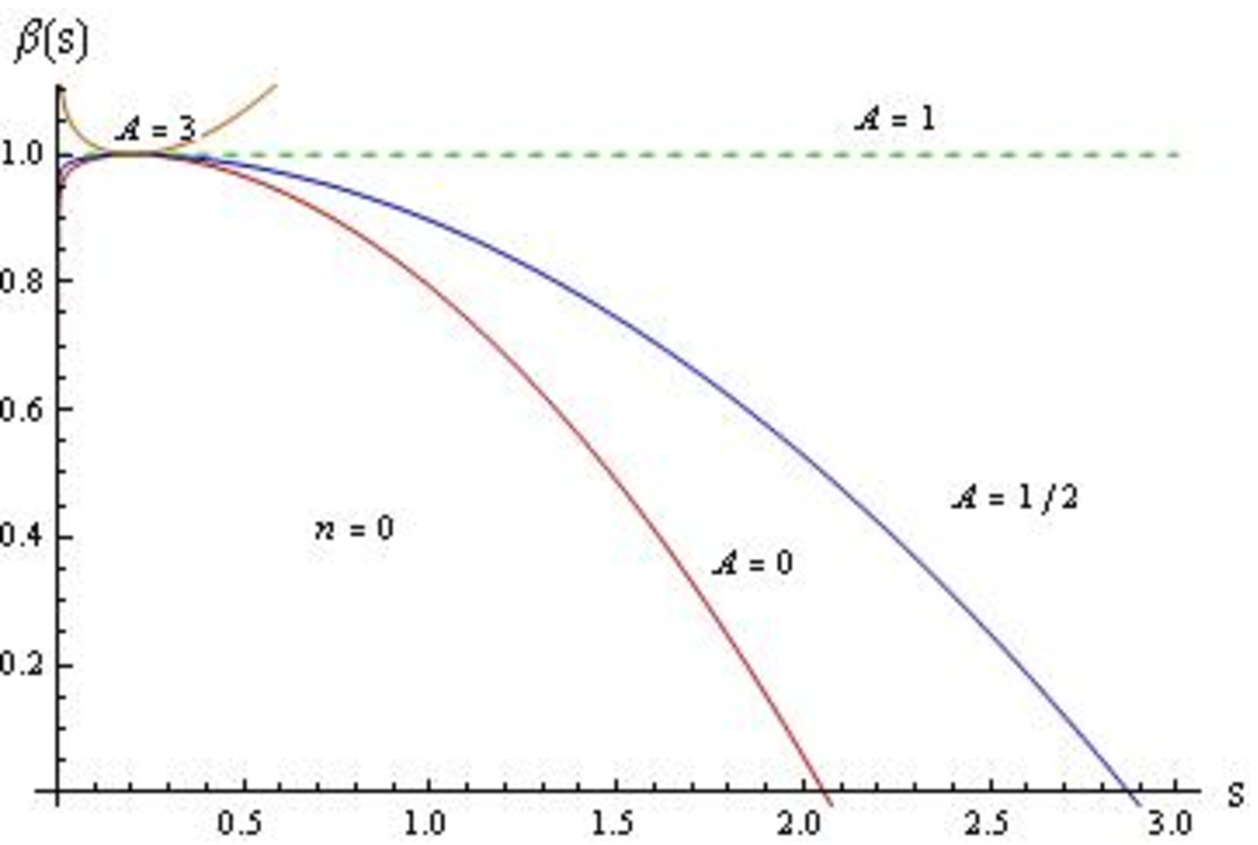,width=.66\linewidth}\\
{\textbf{Figure 1a.} Values of $\beta(s)$ for $n=0$ and
$A=0,1/2,1,3$}.
\end{figure}
\begin{figure}\centering \epsfig{file=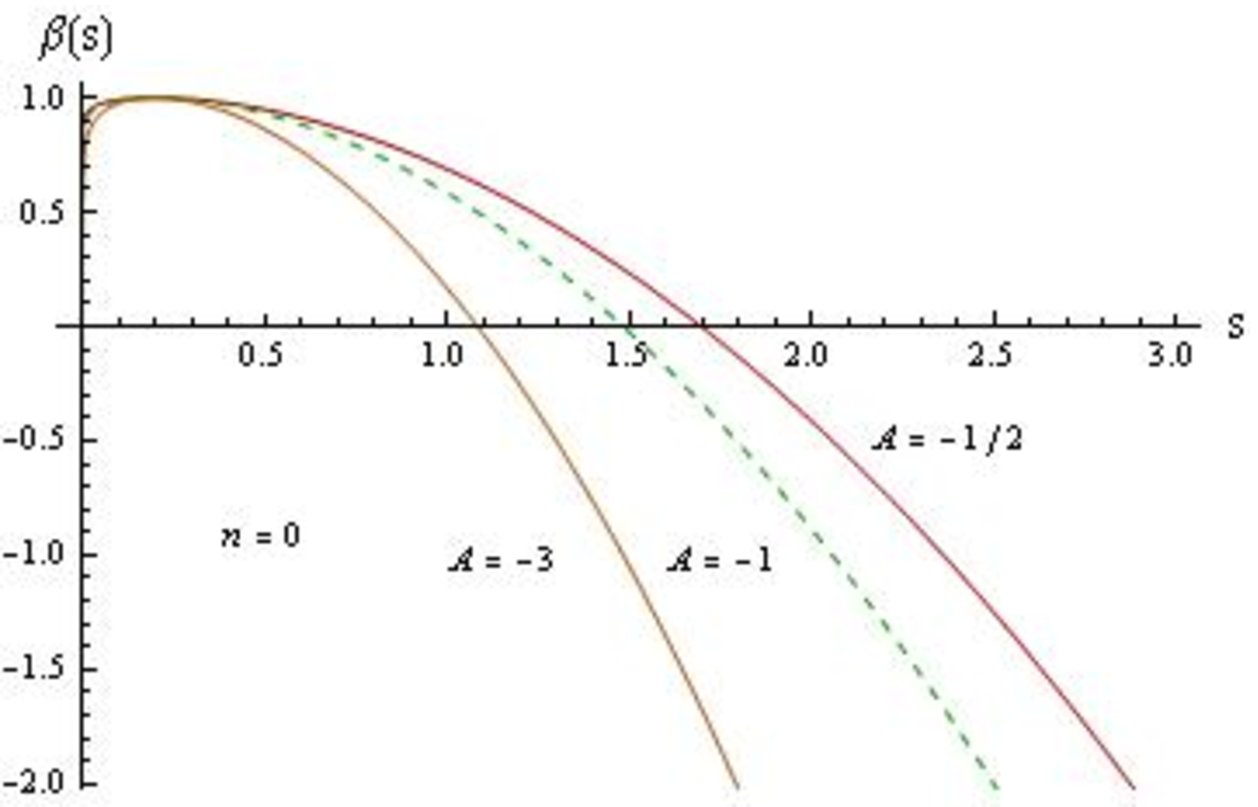,width=.66\linewidth}\\
{\textbf{Figure 1b.} Values of $\beta(s)$ for $n=0$ and
$A=,-1/2,-1,-3$}.
\end{figure}
\begin{figure}\centering \epsfig{file=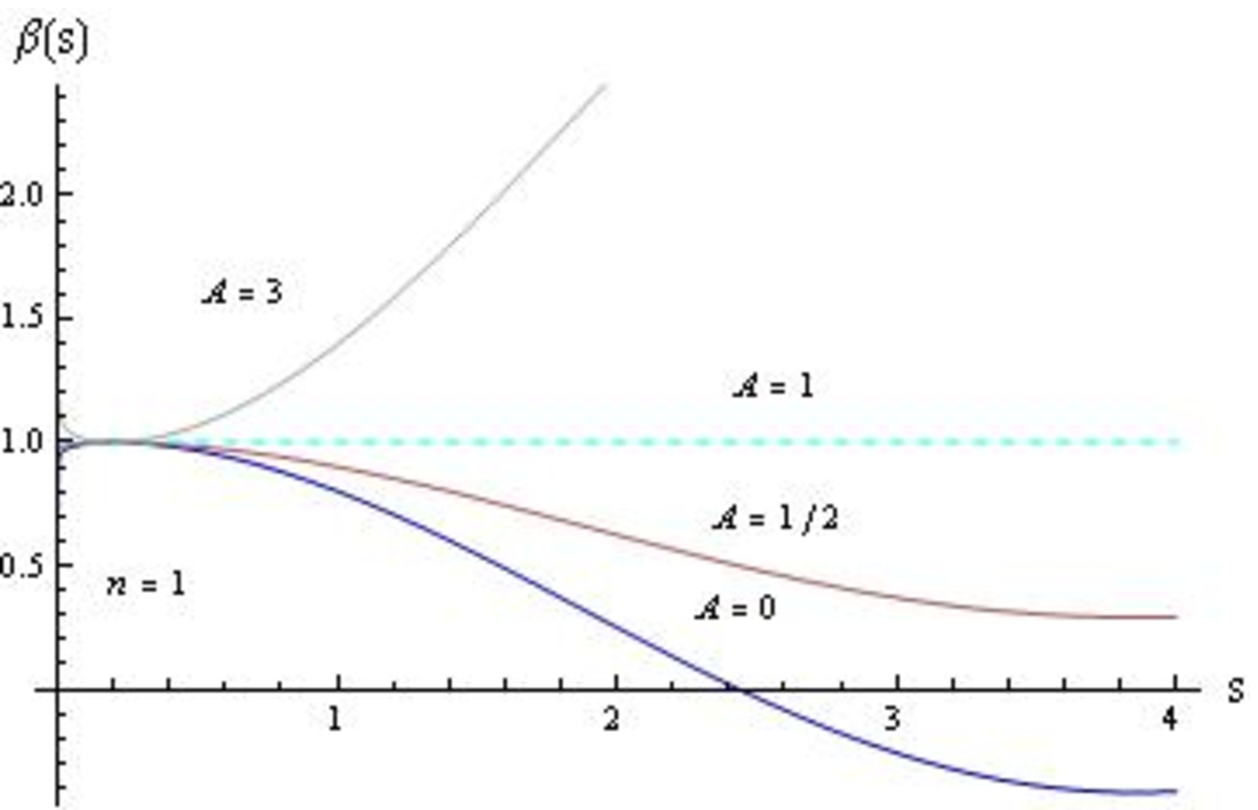,width=.66\linewidth}\\
{\textbf{Figure 2a.} Values of $\beta(s)$ for $n=1$ and
$A=0,1/2,1,3$}.
\end{figure}
\begin{figure}\centering \epsfig{file=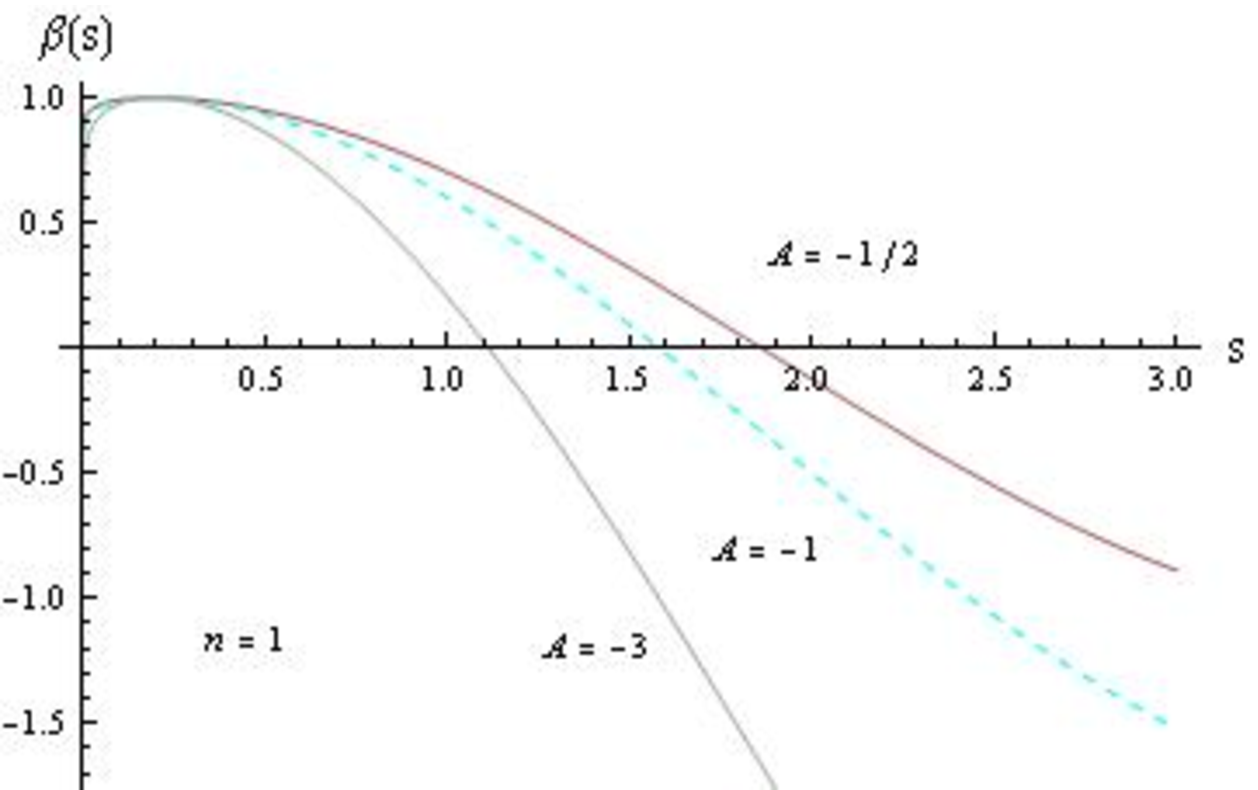,width=.66\linewidth}\\
{\textbf{Figure 2b.} Values of $\beta(s)$ for $n=1$ and
$A=,-1/2,-1,-3$}.
\end{figure}
\begin{figure}\centering \epsfig{file=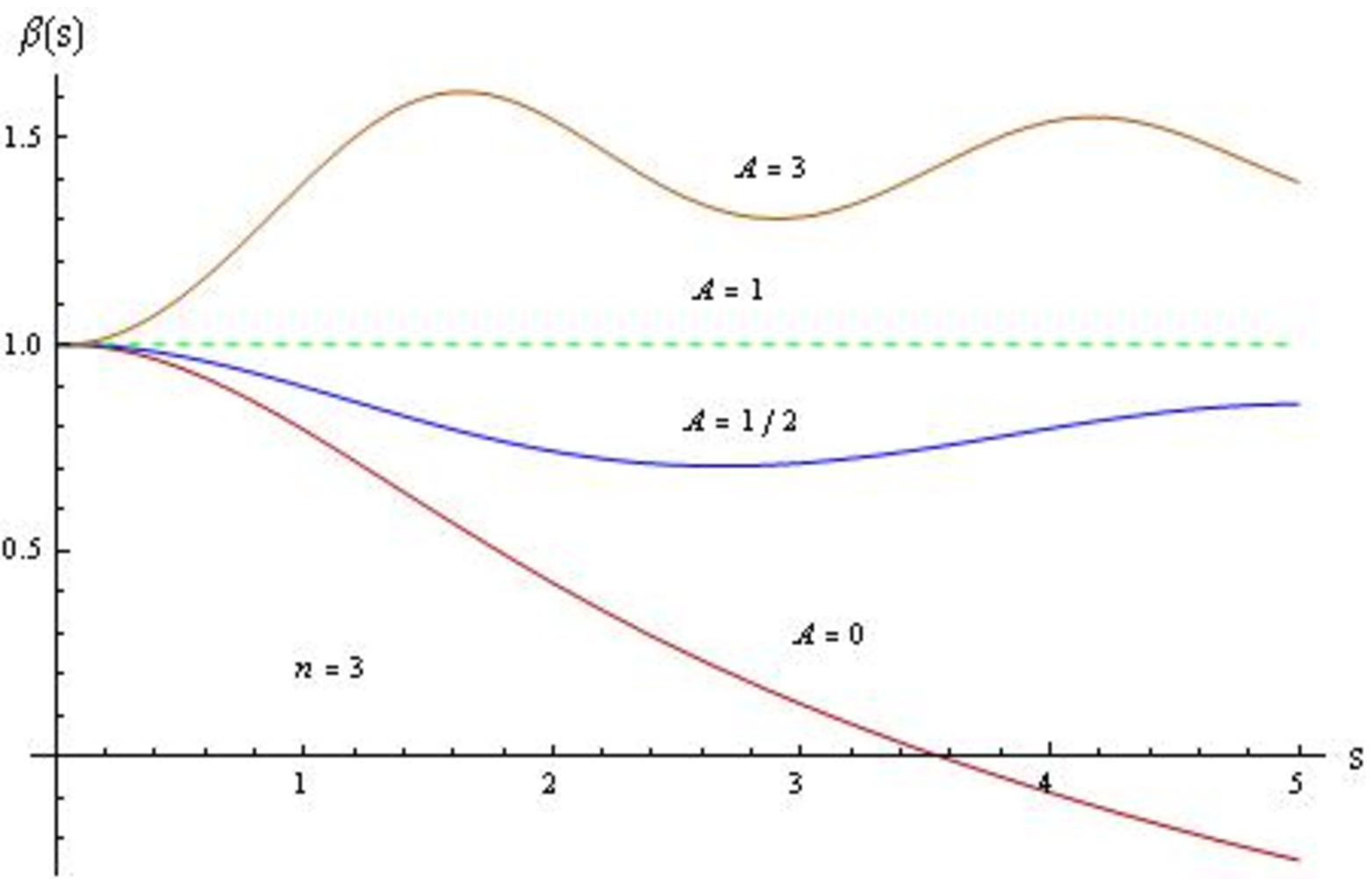,width=.66\linewidth}\\
{\textbf{Figure 3a.} Values of $\beta(s)$ for $n=3$ and
$A=0,1/2,1,3$}.
\end{figure}
\begin{figure}\centering \epsfig{file=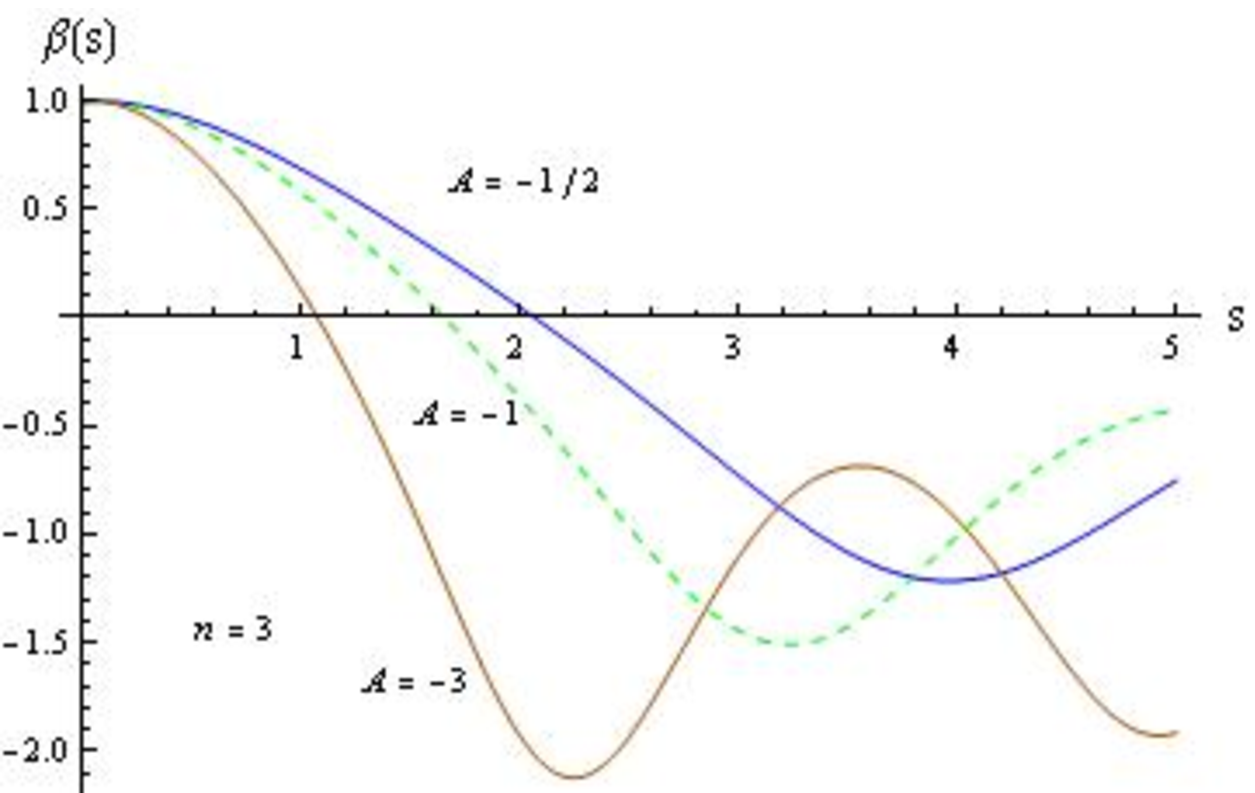,width=.66\linewidth}\\
{\textbf{Figure 3b.} Values of $\beta(s)$ for $n=3$ and
$A=,-1/2,-1,-3$}.
\end{figure}
\begin{figure}\centering \epsfig{file=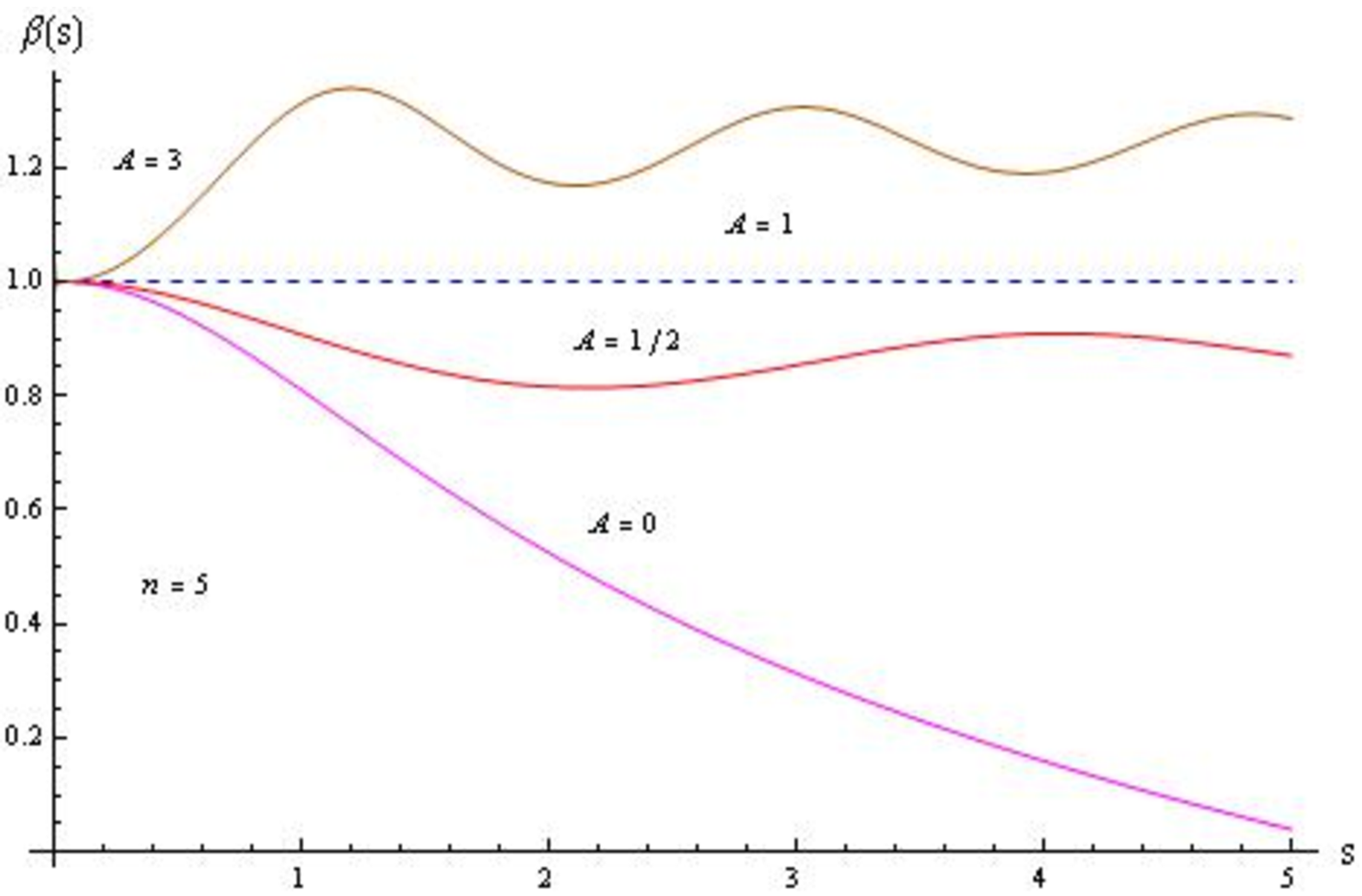,width=.66\linewidth}\\
{\textbf{Figure 4a.} Values of $\beta(s)$ for $n=5$ and
$A=0,1/2,1,3$}.
\end{figure}
\begin{figure}\centering \epsfig{file=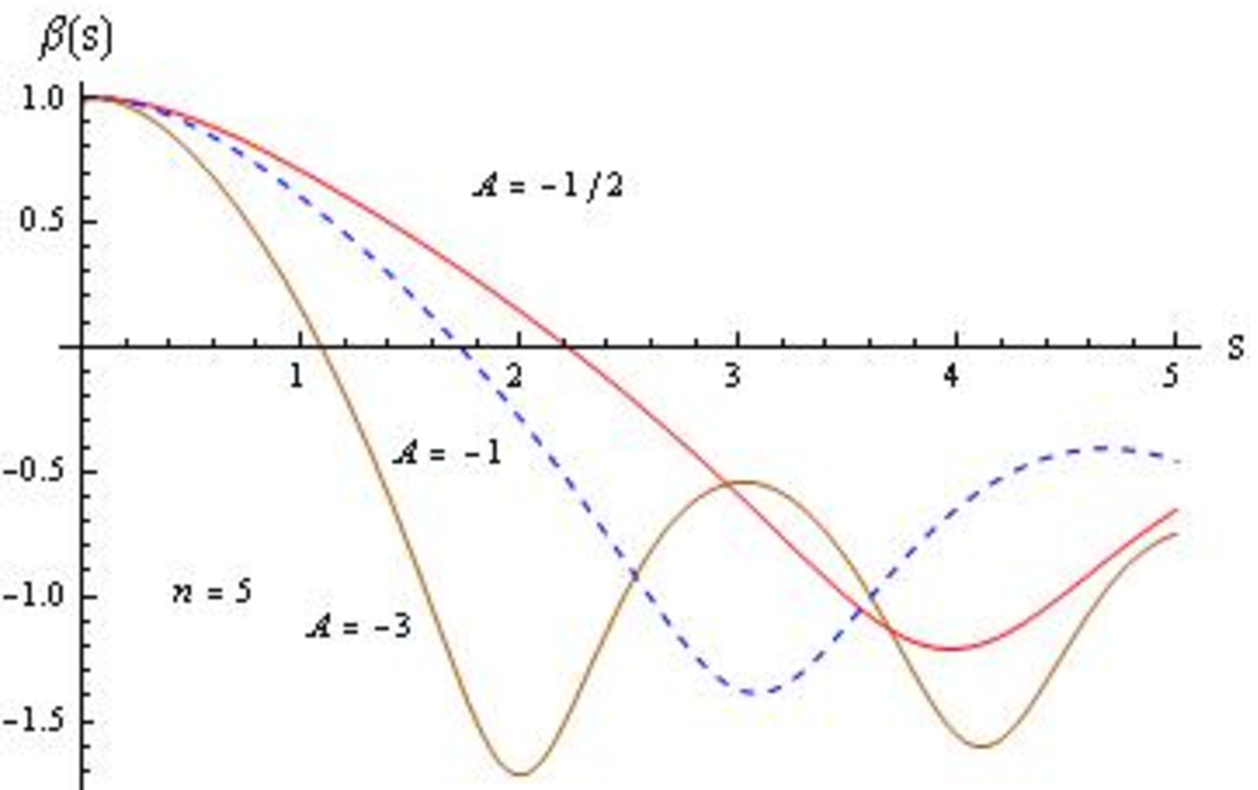,width=.66\linewidth}\\
{\textbf{Figure 4b.} Values of $\beta(s)$ for $n=5$ and
$A=,-1/2,-1,-3$}.
\end{figure}

Figures $\textbf{1a}$ and $\textbf{1b}$ show polytropic models for
$n=0$ with $A=0,\pm1/2,\pm1,\\\pm3$. It is found that the polytropic
filaments for $A\lesssim1/2$ have finite radii, i.e., after finite
values of $s$, the polytropic function $\beta$ becomes zero. For
$A>1/2$, the polytropic models have infinite radii. Figures
$\textbf{2a}$ and $\textbf{2b}$ show that for $n=1$, we have finite
radii models for $A\lesssim0$ and infinite radii models for $A>0$.
Figures $\textbf{3a}$ and $\textbf{3b}$ represent polytropic
filaments for $n=3$ where $A\lesssim0$ represents finite radii while
$A>0$ gives infinite radii models. Figures $\textbf{4a}$ and
$\textbf{4b}$ show models for $n=5$ and the results are almost
similar to the cases $n=1,3$.

It can be noticed that under the constraint
$(m_{0}<<\frac{1}{\tilde{r}})$ and $A=0$, Eq.({\ref{16}) (derived
for SBD gravity) is converted into a non-linear homogenous
Lane-Emden type differential equation which can be used to discuss
polytropic filaments in BD gravity. Thus, numerical results obtained
for $A=0$ and $n=0,1,2,3,5$ can indicate polytropic filaments in BD
gravity. It is mentioned here that the polytropic models described
by GR depend only upon the values of polytropic index $n$. In
spherical case, the polytropic stars for $n<5$ have finite radii
while $n\geqslant5$ provides infinite radii models \cite{24}. In
cylindrical case, there exist models with finite radii for
$0\leqslant n<\infty$ \cite{5}. Since we have derived filamentary
polytopic models present in the universe from small scales to cosmic
scales, so the astrophysical length scale of our evaluated models
are appropriate for both stellar size and cosmic large scales.

\section{Radial Oscillations of Polytropic Filaments}

In this section, we discuss stability as well as normal modes of
radially oscillating filaments about their equilibrium
configuration. For this purpose, we assume that initially the system
is in complete hydrostatic equilibrium. Afterward, the stellar body
starts oscillating (in radial direction) and becomes perturbed
\cite{9,24}.

\subsection{Stability Analysis}

In order to explore radially oscillating filamentary structures for
different matter distribution in SBD gravity, we assume Lagrangian
coordinates $(\mu,t)$, where $\mu$ represents mass per unit length
and $t$ stands for time. Let $r=r(\mu,t)$ be the distance of mass
from the center of cylinder such that $\mu=\pi r^{2}\rho$ satisfies
the following relations \cite{24}
\begin{eqnarray}\label{18}
\frac{d}{d\mu}&=&\frac{1}{2\pi r\rho}\frac{d}{dr},\\\label{18a}
\frac{dr}{d\mu}&=&\frac{1}{2\pi r\rho}.
\end{eqnarray}
To study stability criteria for an oscillating cylindrical filament,
we consider a thin-shell of mass element $d\mu$ (per unit length).
The shell faces a force $F_{g}$ per unit area from the gravitational
field as
\begin{equation}\label{22}
F_{g}=g\rho dr=-\frac{G_{(eff)}\mu}{\pi r^{2}}d\mu-\frac{1}{2\pi
r}\frac{\Lambda_{BD}}{3}d\mu,
\end{equation}
where $g=\frac{G_{(eff)}\mu}{r}+\frac{\Lambda_{BD}}{3}$ represents
gravitational potential due to matter and scalar field. The pressure
gradient forces exert a force $F_{P}$ upon per unit area of the
shell given by
\begin{equation}\label{23}
F_{p}=-\frac{dp}{dr}dr=-\frac{dp}{d\mu}d\mu.
\end{equation}
According to Eqs.(\ref{22}), (\ref{23}) and N approximation
(Newton's second law), the equation of motion for the shell becomes
\begin{equation}\label{23'}
\frac{1}{2\pi
r}\frac{d^{2}r}{dt^{2}}=-\frac{dp}{d\mu}-\frac{G_{(eff)}\mu}{\pi
r^{2}}-\frac{1}{2\pi r}\frac{\Lambda_{BD}}{3}.
\end{equation}

Let us perturb the following quantities adiabatically with time
dependence perturbation $e^{i\omega}$ ($\omega$ indicates frequency
of the oscillations) as
\begin{eqnarray}\label{19}
r(\mu,t)&=&r_{0}(\mu)+r_{0}(\mu)\bar{r}(\mu)e^{i\omega},\\\label{20}
\rho(\mu,t)&=&\rho_{0}(\mu)+\rho_{0}(\mu)\bar{\rho}(\mu)e^{i\omega},\\\label{21}
p(\mu,t)&=&p_{0}(\mu)+p_{0}(\mu)\bar{p}(\mu)e^{i\omega},
\end{eqnarray}
where the quantities with zero subscript indicate unperturbed terms
while the terms with bar represent the perturbed ones. The
perturbations are assumed to be very small such that
$\frac{\bar{r}}{r_{0}},~\frac{\bar{\rho}}{\rho_{0}}$ and
$\frac{\bar{p}}{p_{0}}$ are $<<<1$. After applying the perturbation
scheme to Eq.(\ref{23'}), we obtain
\begin{eqnarray}\label{24}
\frac{dp_{0}}{dr_{0}}&=&g_{0}\rho_{0},\\\label{25}
\frac{d}{d\mu}(p_{0}\bar{p})&=&\frac{\bar{r}}{2\pi
r_{0}}(2g_{0}+r_{0}\omega^{2}),
\end{eqnarray}
where $g_{0}=\frac{G_{(eff)}\mu}{r_{0}}+\frac{\Lambda_{BD}}{3}$ and
we have taken linear contributions of the perturbed terms. Equation
(\ref{24}) represents an unperturbed configuration while
Eq.(\ref{25}) shows perturbed form of equation of motion of the
shell.

Equations (\ref{18}) and (\ref{25}) provide radial dependence of the
perturbed equation of motion as follows
\begin{equation}\label{26}
\frac{d\bar{p}}{dr_{0}}=\frac{\rho_{0}}{p_{0}}
\left[r_{0}\bar{r}\omega^{2}+g_{0}(2\bar{r}+\bar{p})\right],
\end{equation}
while the perturbed configuration of Eq.(\ref{18a}) becomes
\begin{equation}\label{27}
r_{0}\frac{d\bar{r}}{dr_{0}}=-2\bar{r}-\bar{\rho}.
\end{equation}
Since the system is perturbed adiabatically, so the perturbed
density and pressure are related as
\begin{equation}\label{28}
\bar{p}=\Gamma_{(ad)}\bar{\rho}.
\end{equation}
Here $\Gamma_{(ad)}$ is a constant term representing adiabatic
exponent. Using Eqs.(\ref{27}) and (\ref{28}) in (\ref{26}), we
obtain
\begin{equation}\label{29}
\bar{r}''+\left[\frac{3}{r_{0}}-\frac{\rho_{0}g_{0}}{p_{0}}\right]\bar{r}'
+\frac{\rho_{0}}{\Gamma_{ad}P_{0}}(\omega^{2}+
2[1-\Gamma_{(ad)}]\frac{g_{0}}{r_{0}})\bar{r}=0,
\end{equation}
where prime indicates derivative with respect to $r_{0}$. This is a
second order differential equation which represents relative
amplitude $(\bar{r}(r_{0}))$ as a function of depth for a radial
adiabatic oscillation of frequency $\omega$. Equation (\ref{29}) can
be converted into a standard Sturm-Liouville (SL) equation if it is
multiplied by a factor $p_{0}r^{3}_{0}$ as follows
\begin{equation}\label{30}
\left(r^{3}_{0}p_{0}\bar{r}'\right)'
+\frac{r^{3}_{0}\rho_{0}}{\Gamma_{(ad)}}(\omega^{2}+
2[1-\Gamma_{(ad)}]\frac{g_{0}}{r_{0}})\bar{r}=0.
\end{equation}
According to SL theory \cite{24}, the term $\omega^{2}$ behaves as
an eigenvalue for SL problem and there exist infinite number of
eigenvalues $(\omega_{n})$ satisfying the property
$\omega^{2}_{n+1}>\omega^{2}_{n}$. For each eigenvalue, there is a
corresponding eigenfunction $\bar{r}_{n}$ which represents amplitude
of oscillations with $n$ number of nodes in the range
$0<r_{0}<R_{0}$. The lowest-order eigenfunction $\bar{r}_{0}$ shows
no node and is known as the fundamental amplitude.

It is mentioned here that stability, instability as well as
marginally stability of any oscillating model depend upon the
behavior of its frequency. If the frequency of the model is real,
the perturbation is purely oscillatory with constant amplitude
providing a dynamically stable equilibrium, while imaginary values
of the frequency lead to periodic oscillations in which amplitude
increases exponentially in time and the system becomes dynamically
unstable. If the frequency tends to zero, the system becomes
marginally stable (neither stable nor unstable), i.e., the model
will expand and contract with homologous property \cite{24}.

In order to evaluate stability criteria for radial perturbation, we
solve Eq.(\ref{30}) for the fundamental mode with boundary condition
$p_{0}(R_{0})=0$. We insert the eigenfunction $\bar{r}_{0}$ in the
SL equation and integrate over $0<r_{0}<R_{0}$, it follows that
\begin{eqnarray}\nonumber
&&\left[r^{3}_{0}p_{0}\bar{r}'_{0}\right]^{R_{0}}_{0}
+\frac{\omega^{2}_{0}}{\Gamma_{(ad)}}\int^{R_{0}}_{0}r^{3}_{0}\rho_{0}\bar{r}_{0}dr_{0}
+\frac{2-2\Gamma_{(ad)}}{\Gamma_{(ad)}}
\int^{R_{0}}_{0}r^{2}_{0}\rho_{0}g_{0}\bar{r}_{0}dr_{0}=0.\\\label{31}
\end{eqnarray}
Since $\bar{r}'_{0}$ is finite everywhere, so the first term in the
above equation vanishes, yielding
\begin{eqnarray}\label{a}
\omega^{2}_{0}=2(\Gamma_{(ad)}-1)
\frac{\int^{R_{0}}_{0}r^{2}_{0}\rho_{0}g_{0}\bar{r}dr_{0}}
{\int^{R_{0}}_{0}r^{3}_{0}\rho_{0}\bar{r}_{0}dr_{0}}.
\end{eqnarray}
This equation provides the fundamental frequency of the filaments
which leads to the stability criteria of the system. If
$\omega^{2}_{0}>0$ then $\omega^{2}_{n}>\omega^{2}_{0}>0$ for all
nodes $n>0$ and the frequency $(\pm\omega_{n})$ becomes real for all
values of $n$ leading to dynamically stable filaments. Similarly,
$\omega^{2}_{0}<0 \Rightarrow \omega^{2}_{n}<0$ for finite number of
nodes and the frequency becomes imaginary providing unstable
configurations. Otherwise $\omega^{2}_{0}=0$ gives marginally stable
configuration.

Since fundamental amplitude represents no node, so it has the same
sign all over the cylindrical filament and hence both the integrals
in Eq.(\ref{a}) have the same sign. Thus, the stability criteria
depends upon $ sign~\omega_{0}^{2}=sign~2(\Gamma_{(ad)}-1)$. If
\begin{equation}\label{32}
\Gamma_{(ad)}>1,
\end{equation}
$\omega^{2}_{0}>0$ and the corresponding model becomes stable. For
\begin{equation}\label{33}
\Gamma_{(ad)}=1,
\end{equation}
$\omega^{2}_{n}=0$ and the system is marginally stable. When
\begin{equation}\label{34}
\Gamma_{(ad)}<1,
\end{equation}
$\omega^{2}_{0}<0$ the system becomes dynamically unstable.

Thus, the criteria of stability for the filamentary structures in
the N limit of SBD gravity depends only upon stiffness of the fluid
$(\Gamma_{(ad)})$. The stability conditions are independent from the
behaviors of dynamical variables related to metric, coupling
constant $(\omega_{BD})$, scalar field and potential of the scalar
field.

It is mentioned here that in GR, the stability of spherical as well
as cylindrical polytropic models depends only upon stiffness
parameter. In the spherical case, $\Gamma_{(ad)}<\frac{4}{3}$ leads
to unstable model while cylindrically symmetric filaments remain
stable for $\Gamma_{(ad)}>1$\cite{9,b}. In the case of BD gravity,
the spherically symmetric polytropic model remains unstable for
$\Gamma_{(ad)}>\frac{4}{3}$ \cite{c} while cylindrically symmetric
polytropic models are not discussed. Thus, our obtained results are
consistent with GR in N regime.

\subsection{Normal Modes of Radial Oscillations}

The behavior of normal modes for polytropic filaments can be
described by using polytropic quantities in the SL problem. We use
Eqs.(\ref{10}), (\ref{13}), (\ref{15}) in (\ref{30}), it follows
that
\begin{eqnarray}\nonumber
&&\frac{d\bar{r}^{2}}{ds^{2}}+\left[\frac{3}{s}
+\frac{n+1}{\beta}\frac{d\beta}{ds}-\frac{\Lambda_{BD}}{3}\right]\frac{d\bar{r}}{ds}
-\left[b^{2}\frac{n+1}{\Gamma_{(ad)}h_{00(c)}}\omega^{2}\right.\\\label{35}
&&\left.+2(1-\Gamma_{(ad)})\frac{n+1}{\Gamma_{(ad)}s}
\frac{d\beta}{ds}\right]\frac{\bar{r}}{\beta}=0.
\end{eqnarray}
In order to solve this equation for $\bar{r}$, the required boundary
conditions are as follows: at the center of cylinder, we have
$\left(\frac{d\bar{r}}{ds}\right)_{s=0}=0$, at the boundary of
polytrope $s=S$, we have $\frac{p_{0}}{\rho_{0}}<<<1$ and the values
of $\bar{r}$ remain finite. Thus, Eqs.(\ref{26}) and (\ref{27})
yield
\begin{equation}\nonumber
[\omega^{2}bS+g_{0}(2-2\Gamma_{(ad)})]\bar{r}(S)
-g_{0}\Gamma_{(ad)}(\frac{d\bar{r}}{ds})_{s=S}=0.
\end{equation}
Equation (\ref{35}) along with boundary conditions provide different
patterns of normal modes of oscillating polytropic filaments in SBD
theory. It is mentioned here that the behavior of these normal modes
depends upon the frequency, central potential, polytropic index,
adiabatic exponent and gravitational potential term due to massive
scalar field $(g_{0})$, i.e., for different values of these
parameters, we have different modes of radial oscillations of
polytropic filaments.

\section{Final Remarks}

This paper investigates cylindrically symmetric filamentary
structures in N limit  of SBD gravity. We have formulated a
generalized form of Lane-Emden equation in N approximations and
obtained polytropic filament models analytically for $n=0,1$ as well
as numerically for $n=0,1,3,5$. We have found that the behavior of
these models depend upon the values of polytropic index as well as
the cosmological constant term $A$ (due to the scalar field). For
$n=0$, the models have finite radii for $A\lesssim1/2$, otherwise
they have infinite radii. For $n=1,3,5$, we have approximated that
$A\lesssim0$ represents finite radii polytropic filaments while
$A>0$ gives infinite radii models. We have also found that within
$m_{0}<<\frac{1}{\tilde{r}}$ and $A=0$, the models defined for $A=0$
and $n=0,1,2,3,5$ are same for BD gravity.

In order to study fragmentation of filamentary structures, we have
investigated stability of radial oscillations of polytropic
filaments. It is found that the stability criteria of cylindrical
filaments in SBD gravity depends only upon adiabatic index
($\Gamma_{ad}$). Generally in the weak field approximation of GR and
BD gravity, the stability criteria depends upon the adiabatic index
as well as on the dynamical variables related to matter and scalar
field distributions. But in the polytropic case, the adiabatic index
is responsible for the stability of the function. Finally, we have
discussed possible normal modes of radial oscillations of polytropic
filaments. It turns out that different values of parameters lead to
different modes of radially oscillating filaments. It is found that
weak field approximation of SBD gravity is consistent with
observations for all arbitrary values $\omega_{BD}$ and hence our
obtain results are valid for all arbitrary values of $\omega_{BD}$.
It is interesting to mention here that our results provide
generalized form of cylindrical filament polytropic models of GR
theory.

\end{document}